\input epsf.sty
\documentclass[aps,prl,twocolumn,superscriptaddress,showpacs]{revtex4}
\usepackage{graphicx}
\begin{document}
\title{
Spin Accumulation in the Electron Transport
with Rashba Interaction
}
\author{Yi-Ying Chin}
\affiliation{
Department of Physics, National Tsing Hua University,
Hsinchu 30043, Taiwan}
\author{Jui-Yu Chiu}
\affiliation{Dept. of Physics, University of California-San Diego, La Jolla,
California 92093-0319, USA}
\author{Ming-Che Chang}
\affiliation{Dept. of Physics, National Normal University,
Taipei, Taiwan}
\author{Chung-Yu Mou}
\affiliation{
 Department of Physics, National Tsing Hua University,
Hsinchu 30043, Taiwan}
\affiliation{Physics Division, National Center for Theoretical
Sciences, P.O.Box 2-131, Hsinchu, Taiwan}
\date{\today}
\begin{abstract}
The non-equilibrium transportation of two-dimensional electrons
through a narrow channel
is investigated under the influence of the Rashba interaction.
By introducing suitable lifetime in the Green's function,
the average spin values can be calculated from the
ballistic regime to the diffusive regime. It is shown that
the spin accumulation is a combined effect of the spin current
and disorders. In the diffusive regime, disorders offer a mechanism
to stop the spin current and generate the spin accumulation.
In the ballistic regime, spins are
more spread out and do not have definite signs.
Further consideration
indicates that the inclusion of ferromagnetic spin-spin 
interaction increases the
spin accumulation near the edge.
\end{abstract}

\pacs{72.25.-b, 72.25.Dc, 73.40.-c} \maketitle

Generation and transfer of spins, preferably without using external magnetic
field or magnetic materials, is an important issue in spintronics\cite
{zutic04}. In semiconductors, through the intrinsic spin-orbit interactions
related to the Dresselhaus\cite{dresselhaus} or the Rashba\cite{bychkov84}
mechanism, it is possible to generate, or even manipulate spins using only
an electric field. Recently, it is further proposed that the
spin Hall current can be generated in bulk semiconductors\cite{murakami} and
semiconductor heterojunctions\cite{sinova} because of the spin-orbit
interactions. Unlike an earlier proposed spin Hall effect involving
impurities\cite{hirsch}, these two studies demonstrated that the spin Hall
effect is possible in pure samples without any impurities. Experimentally, 
the spin Hall effect has been observed in a number of
systems\cite{kato}. The key observation is the accumulation of
opposite spin polarizations on two sides of the sample. 
While it is natural
to associate the spin-accumulation with the spin Hall effect, 
for the intrinsic case, 
a satisfying theory of spin accumulation still does not exist.

In this paper, by introducing suitable lifetime in the Green's
function, the spin accumulation will be calculated from the ballistic regime
to the diffusive regime. The origin of spin
accumulation will be clarified.

We start by considering a two-dimensional electron system confined in a
channel 
\begin{equation}
H=-\,\frac{\hbar ^{2}}{2m}\bigtriangledown ^{2}+\frac{\alpha }{\hbar }\,\hat{
z}\cdot \lbrack \,\hat{\sigma}\times \hat{p}\,]+V_{C}+V_{I}.  \label{eq1}
\end{equation}
Here $\alpha$ characterizes the Rashba iteraction, 
$\hat{\sigma}$ are Pauli matrices and $V_{C}$ is the hard-wall
potential that confines the electron to $-W/2<y<W/2$ . $V_{I}$ is the
potential due to point disorders, $V_{I}(\mathbf{r}
)=\sum_{i}\upsilon _{0}\delta (\mathbf{r-R}_{i})$\cite{Mahan}. We shall use $
\overline{\cdots }$ to denote averaging over disorders and $n_{i}$ to
denote the concentration of impurities. Since $H$ possesses the
time-reversal symmetry, the Kramer's theorem implies that each 
eigenstate is doubly-degenerate.
Physically, the  degeneracy represents two opposite propagating
directions along $x$ axis. Therefore, we shall denote the Kramer degeneracy
by $\lambda =\pm $, representing propagating along $\pm x$ respectively.

To find the spin-accumulation $\overline{\langle \,S_{z}(\mathbf{r}
)\,\rangle }$ ($\equiv s_{z}(\mathbf{r})$), we first note that because $
\sigma _{y}H(x,y)\sigma _{y}=H(x,-y)$, for each eigenstate $\psi
_{n\lambda }(x,y)$, one obtains $\sigma _{y}\psi _{n\lambda }(x,y)=\psi
_{n\lambda }(x,-y)$ and thus it follows that $s_{z}(\mathbf{r})$
is antisymmetric in $y$. This simple argument, however, can not
assign a definite sign for $s_{z}(\mathbf{r})$. 
Numerics is then necessary to pin down the sign. It is important to note
that for each energy, $s_{z}(\mathbf{r})$ have
opposite signs for opposite propagating directions. As a result,
spin-accumulation can only happen when the system is not in equilibrium,
i.e., there must be a biased potential between two electrodes. 
Therefore, one has  $s_{z}(\mathbf{r})$ =$\sum_{n}\overline{
\langle \,\psi _{n+}^{\alpha }(\mathbf{r})\mid S_{z}^{\alpha \beta }\mid
\psi _{n+}^{\beta }(\mathbf{r})\,\rangle }$ , where $\psi _{n+}^{\beta }(
\mathbf{r})$ denotes eigenstates with $\lambda =+$ and the energy $
E_{n}$ is restricted to the regime $eV_{1}\leq E_{n}\leq eV_{2}$ with $V_{1}$
and $V_{2}$ being the electric potentials of electrodes. By inserting $
\int_{eV_{1}}^{eV2}dE\delta (E-E_{n})$ and using the identity $\mathrm{Im}
[1/(E-E_{n}+i\epsilon )]=-i\pi \delta (E-E_{n})$, 
we find
\begin{equation}
\overline{\langle \,S_{z}(\mathbf{r})\,\rangle }=\,\int_{eV_{1}}^{eV2}dE
\left\{ \frac{-1}{\pi }\mathrm{Im}Tr[S_{z}\overline{G_{+}(E,\mathbf{r},
\mathbf{r})}]\right\} ,  \label{eq2}
\end{equation}
where $G_{+}(E,\mathbf{r},\mathbf{r})$ is the retarded Green's function in
the\ $\lambda =+$ channel and \ $Tr$ is the trace over the spin space.
In the Born approximation, 
$\overline{G_{+}^{\alpha \beta }}(E,\mathbf{r},\mathbf{r}
)=\sum_{n}[\phi _{n+}^{\alpha }(\mathbf{r})]^{\ast }\phi _{n+}^{\beta }(
\mathbf{r})/[E-E_{n}+i\gamma N_{+}(E,\mathbf{r})+i\epsilon ]$ with $\gamma
=n_{i}\upsilon _{0}^{2}$, $\phi _{n+}(\mathbf{r})$ being the energy
eigenstate and $N_{+}(E,\mathbf{r)}$ being the local density of states in
the absence of disorders\cite{Mahan}. 
Because $k_{x}$
is a good quantum number in the absence of disorders, 
$\lambda =+$ corresponds to positive $k_x$. For a given $k_x$,
$\phi _{n+}(\mathbf{r})$ is a linear combination of four
waves with $k_y = \pm k_{y}^{\pm }$, 
where $E\equiv \hbar ^{2}k_{\pm }^{2}/2m\pm \alpha k_{\pm }$ with $k_{\pm
}^{2}=$ $k_{x}^{2}+(k_{y}^{\pm })^{2}$. 
The hard-wall boundary condition yields the following equation 
and further 
selects 
$k_{y}^{\pm }$
\begin{eqnarray}
&&1+e^{2i(k_{y}^{+}+k_{y}^{-})W}-4e^{i(k_{y}^{+}+k_{y}^{-})W}\frac{
k_{y}^{+}k_{y}^{-}}{k_{+}k_{-}+k_{x}^{2}+k_{y}^{+}k_{y}^{-}}  \nonumber \\
&&-(e^{ik_{y}^{+}W}+e^{ik_{y}^{-}W})\frac{
k_{+}k_{-}+k_{x}^{2}-k_{y}^{+}k_{y}^{-}}{
k_{+}k_{-}+k_{x}^{2}+k_{y}^{+}k_{y}^{-}}=0.  \label{eq3}
\end{eqnarray}
Once $k_{y}^{\pm }$ are found, $\phi
_{n+}$ and thus $\overline{G_{+}^{\alpha \beta }}(E,\mathbf{r},\mathbf{r})$
are obtained. Clearly, $k_{y}^{\pm }$ determine how $s_{z}(\mathbf{r})$
oscillates for each state and in general, there is no obvious spin
accumulation for each state. This is similar to the charge Hall effect in which
there is no charge accumulation for single particle states and one needs
Coulomb interaction to stop charges and generate charge accumulation. 
Fig. 1 shows the
numerical results of $s_{z}(\mathbf{r})$ for different $\gamma $. By setting 
$\gamma \rightarrow 0$, one reaches the ballistic limit where 
spins are more spread out, and furthermore, as shown in the inset of Fig. 1,
depending on parameters, $s_{z}(\mathbf{r})$ may even switch the sign.
Nonetheless, as one turns on disorders,
$s_{z}(\mathbf{r})$ begins to accumulate near the edge and always
switches to be negative on the left hand side (LHS) and positive on the
right hand side (RHS). Obviously, disorder is mechanism for generating
spin accumulation. The reason can be found by exploring the continuity
equation $\partial s_{z}(\mathbf{r},t)/\partial t=-s_{z}(\mathbf{r},t)/\tau -
\mathbf{\nabla }\cdot \mathbf{J}_{z}(\mathbf{r},t)$ where $\tau $ is the
effective diffusion time due to disorders and $\mathbf{J}_{z}$ is the spin
current. In the steady state, one obtains $s_{z}(\mathbf{r})=-\tau \mathbf{
\nabla }\cdot \mathbf{J}_{z}(\mathbf{r})$, which implies that if $\mathbf{J}
_{z}$ flows from right to left, $s_{z}(\mathbf{r})$ is negative on LHS and
positive on RHS, in consistent with numerical results. 
Thus the spin
diffusion generates the spin
accumulation that stops the spin current.

In addition to the spin diffusion, the spin-spin interaction, $J\sum \mathbf{
\sigma}_{i}\cdot \mathbf{\sigma}_{j}$, further enhances the spin accumulation. Fig. 2
shows our numerical results based on the self-consistent mean-field theory .
Clearly, 
ferromagnetic coupling amplifies the oscillation of the spin density and
increases the accumulation. 
For antiferromagnetic coupling,  the oscillation has a
shorter length scale because neighboring spins tend to be antiparallel.

In conclusion, we have investigated the spin accumulation from the ballistic
limit to the diffusive regime. The origin of the spin accumulation is
clarified. This research was supported by NSC of Taiwan.

\begin{figure}
\rotatebox{-90}{\includegraphics*[width=60mm]{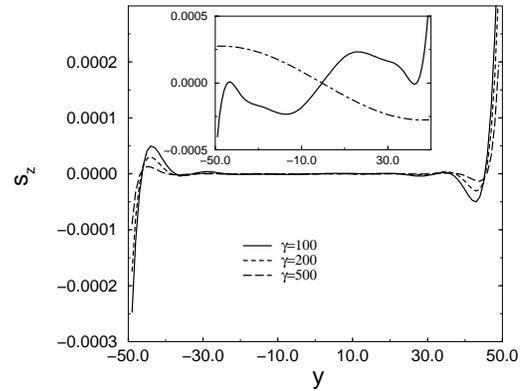}}
\caption{Spin accumulation for different $\gamma$. Here $eV_1=50$,
$eV_2 =0$, and $k_0 W=3$ with $k_0=2m \alpha/\hbar ^{2}$
and energy being in unit of $k_0 \alpha$.
Inset: Solid line: $\gamma=1$ and other parameters
are the same, dash line: $\gamma=0.01$, $eV_1=30$,
$eV_2 =0$, and $k_0 W=1$} \label{fig-1}
\end{figure}

\begin{figure}
\rotatebox{-90}{\includegraphics*[width=60mm]{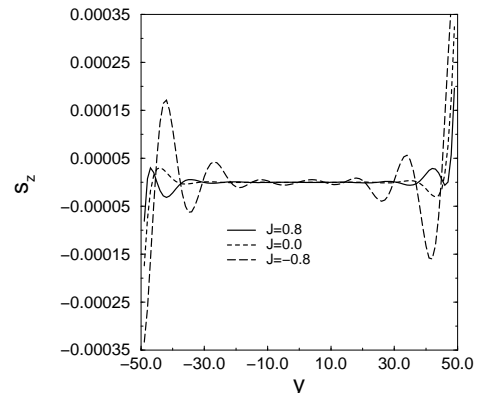}}
\caption{Effects of ferromagnetic coupling on spin accumulation with the same
parameters used in Fig.~1. Here
$J$ is in unit $meV$.
} \label{fig-2}
\end{figure}



\appendix


\begin{thebibliography}{99}
\bibitem{zutic04}  Igor Zutic, Jaroslav Fabian, S. Das Sarma, \textit{Rev.
Mod. Phys.} 76 (2004), p. 323.

\bibitem{dresselhaus}  G. Dresselhaus, \textit{Phys. Rev.} 100 (1955), p.
580.

\bibitem{bychkov84}  E.~I. Rashba, \textit{Sov. Phys. Solid State} 2 (1960),
p. 1224 .

\bibitem{murakami}  S. Murakami, N. Nagaosa, and S.~C. Zhang, \textit{Science} 
01 (2003), p. 1348 ; \textit{
Phys. Rev. B} 69 (2004), 235206.

\bibitem{sinova}  Jairo Sinova, Dimitrie Culcer, Q. Niu, N.~A. Sinitsyn, T.
Jungwirth, and A.~H. MacDonald, \textit{Phys. Rev. Lett.} 92(2004), 126603.

\bibitem{hirsch}  J.E. Hirsch, \textit{Phys. Rev. Lett.} 83 (1999), p. 1834.

\bibitem{kato}
Y.K. Kato, R.C. Myers, A.C. Gossard, and D.D. Awschalom, Science
{\bf 306}, 1910 (2004); J. Wunderlich, B. Kaestner, J. Sinova, and 
T. Jungwirth, 
Phys. Rev. Lett. {\bf 94}, 047204 (2005). 




\bibitem{Mahan}  G. D. Mahan, Many-particle Physics, p264, 2nd Edition,
Plenum Press, New York (1990).
\end{thebibliography}
\end{document}